\documentclass{ws-p8-50x6-00}
\def\Dirac#1{#1\hskip-6pt/}
\def\dd{\Dirac\partial}
\def\dD{\nabla\hskip-8pt/}
\def\dpp{\Dirac p}

\begin{document}

\title{chiral symmetry breaking and color superconductivity in the
instanton picture}

\author{Gregory W. Carter}

\address{The Niels Bohr Institute, Blegdamsvej 17, 2100 Copenhagen, Denmark}

\author{Dmitri Diakonov}

\address{NORDITA, Blegdamsvej 17, 2100 Copenhagen, Denmark}


\maketitle

\abstracts{
The instanton approach to spontaneous chiral symmetry breaking is reviewed,
with emphasis on the connection to chiral random matrix theory.  We extend
the approach to discuss the finite density, zero-temperature behaviour of
quark matter.
Since the instanton-induced interactions are
attractive in both $\bar{q}q$ and $qq$ channels, a competition ensues
between phases of matter with condensation in either or both.
It results in chiral symmetry restoration due to the onset of diquark
condensation, a `colour superconductor', at finite density.
}

\section{Why Use Instantons?}

The idea that the QCD partition function is dominated
by instanton fluctuations of the gluon field, with quantum
oscillations about them, has successfully confronted many 
facts we know about the zero-temperature, zero-density hadronic world
(for reviews see refs. \cite{D1,SS}).  Instantons have been reliably
identified in lattice simulations (for a review see ref. \cite{vB}).
From the theory side, the instanton vacuum constructed
from the Feynman variational principle\cite{DP1}, gives
an example of how the necessary `transmutation of dimensions' can
actually happen in QCD, meaning that all dimensional quantities
can be expressed through the QCD scale parameter, $\Lambda_{QCD}$.

\section{Instanton Partition Function}

The strategy of the instanton approach is to replace the
partition function of full QCD with a sum over instantons and quantum
fluctuations about them.  In particular, the result is an integration
over collective coordinates of instantons.
A finite-density extension of this model is presumed valid for quark
chemical potentials less than approximately 600 MeV, a scale set by the
instantons themselves.

The QCD partition function is approximated in two steps. First, the gluon
degrees of freedom are replaced with a sum over an ensemble of
instantons ($I$'s) and anti-instantons ($\bar I$'s) with
quantum fluctuations above the non-perturbative vacuum. Explicitly, we
write
\begin{equation}
A_{\mu} = \sum A^I_{\mu}(\xi) + \sum
A^{\bar{I}}_{\mu}(\xi) + B_{\mu}\,,
\end{equation}
where $B_{\mu}$ are quantum fluctuations above the instanton vacuum and
$\xi$ are the collective instanton coordinates comprised of positions,
sizes, and colour orientations.  With this ansatz, the partition
function, including the quark chemical potential, is
\begin{equation}
{\cal Z} = \sum\frac{1}{N_+! N_-!} \int \!
d\xi\; {\cal D} B_{\mu}\; e^{-U_{inst}(\xi)} \frac{ {\rm
det}(i\dD+im-i\mu\gamma_4){\rm det} (i\dd+iM)}{{\rm det}(i\dD+iM){\rm
det}(i\dd+im)}.
\end{equation}
Here $m$ is a current quark mass, whereas $M$ is a Pauli-Villars
regulator mass.  This expression provides a lower bound on the
partition function\cite{DP1}.

Next, the quark determinant is partitioned into high and low momentum
contributions. These regimes are divided by an arbitrary mass parameter 
$M_1$, which is chosen large enough such that Det$_{high}$ is 
factorizable and small enough so that Det$_{low}$ is saturated by the 
zero modes\cite{DP2}. With it, we partition as
\begin{eqnarray}
{\rm Det} &=& {\rm Det}_{low}\cdot{\rm Det}_{high} \nonumber\\
{\rm Det}_{low} &=& \frac{ {\rm det}(i\dD+im-i\mu\gamma_4){\rm det}(i\dd+iM_1)}
{{\rm det}(i\dD+iM_1){\rm det}(i\dd+im)} \nonumber\\
{\rm Det}_{high} &=& \frac{ {\rm det}(i\dD+iM_1){\rm det}(i\dd+iM)}
{{\rm det}(i\dd+iM_1){\rm det}(i\dD+iM)}
\end{eqnarray}
If we consider the fields $B_{\mu}$ to be perturbative and subsequentially
disregard their contributions, we may perform the averaging over the
background gauge fields represented by instantons.  For the low momentum part
one gets the form\cite{DP2}:
\begin{equation}
{\cal Z}_{low} = \sum_{N_\pm}\frac{e^{i\theta(N_+-N_-)}}{N_+! N_-!}
\int\!\prod_{I\bar I}
d^4z_I d\rho_I d{\cal O}_I\; d(\rho_I) \prod_f^{N_f} 
\left(\begin{array}{cc}
im_f & T_{I\bar I} \\ T_{\bar I I} & im_f \end{array}\right)
\label{partfu}
\end{equation}
The $T$ matrix is of size $N_+\times N_-$, where $N_\pm$ are the 
number of $I$'s and $\bar I$'s, and is comprised of the overlaps of 
would-be zero modes,
\begin{equation}
T_{I\bar I} = \int\!d^4x\;\tilde{\Phi}_0(x-z_I,\rho_I,{\cal O}_I;\mu)
(i\dd-i\mu\gamma_4)\Phi_0(x-z_{\bar I},\rho_{\bar I},{\cal O}_{\bar I};\mu)
\,.
\end{equation}
These integrals are not only functions of the instanton coordinates, but have
parametric dependence on chemical potential and temperature as well.

\section{Chiral Symmetry Breaking by Instantons}

The instanton partition function may be analyzed in three ways, each
conceptually quite different but leading to equivalent results. In 
each case, the main result is spontaneous breaking of chiral symmetry 
in the vacuum. This phenomena can be interpreted as a delocalization 
of the ``would-be'' zero modes, induced by the background instantons, 
resulting from quarks hopping between them\cite{DP2}. It was first 
noticed in ref. \cite{DP3} that there is a far reaching analogy 
between chiral symmetry breaking in QCD and the problem of electrons
in condensed matter systems with random impurities. The acquisition
of a dynamical (sometimes called constituent) mass by a quark is
fully analogous to the appearance in the Green function of an 
electron in a metal of a finite relaxation time (but in our case
this time depends on the momentum). The appearance of the massless
pole in the pseudoscalar channel corresponding to the Goldstone pion
is analogous to the formation of a diffusion mode in the 
density-density correlation function.  For the recent development of these and
related ideas, see Refs. \cite{JNPZ} and references therein.

\subsection{Random Matrices Approach}

Eq.\,(\ref{partfu}) is a particular (and historically first)
example of a matrix approach to chiral symmetry breaking.
The matrices here are made of the overlaps $T_{I\bar I}$
of the would-be fermion zero modes in the background of instantons
whose positions, sizes and orientations are random. To make a quick
estimate of the spectral density of the Dirac operator, $\nu(\lambda)$, one can
use the cluster decomposition in the elements of the $T_{I\bar I}$
matrix\cite{DP2}:
\begin{equation}
\nu(\lambda) = \overline{\sum_n \delta(\lambda-\lambda_n)} =
\int\frac{ds}{2\pi}\;e^{-is\lambda}\;\overline{{\rm Tr}\;e^{isT}}\,,
\end{equation}
where the averaged trace over an ensemble of matrices is expanded as
\begin{equation}
\overline{{\rm Tr}\;e^{isT}} = N{\rm exp}\left\{ is\frac{1}{N}\overline{
{\rm Tr}\;T} + \frac{(is)^2}{2}\left[\frac{1}{N}\overline{{\rm Tr}\;T^2} -
\frac{1}{N^2}\overline{{\rm Tr}\;T}^2\right] + \ldots\right\}\,.
\label{cummul}
\end{equation}
The first term vanishes upon averaging. From the leading cummulant 
one has
\begin{equation}
\nu(\lambda) = \frac{N}{\sqrt{2\pi\kappa^2}}\;e^{-\frac{\lambda^2}
{2\kappa^2}}\,,
\end{equation}
where $N$ is the total number of $I$'s and $\bar I$'s and the parameter
$\kappa$ is the average overlap of zero modes:
\begin{equation}
\kappa^2 = \frac{1}{N}\overline{{\rm Tr}\left[T_{I\bar I} T_{\bar I I}\right]}
={\rm const}\cdot\frac{N}{V}\bar\rho^2\,,
\label{defkap}
\end{equation}
with $\bar\rho^2$ being the average size of instantons obtained from the
size distribution $d(\rho_I)$ of eq.\,(\ref{partfu}). 

Using the first cummulant in eq.\,(\ref{cummul}) is, however, incomplete.  
Resumming all contributions yields the Wigner semicircle spectrum:
\begin{equation} 
\nu(\lambda) = 
\frac{N}{\pi\kappa}\sqrt{1-\frac{\lambda^2}{4\kappa^2}}\,.  
\label{Wigner}
\end{equation}

The chiral condensate is obtained directly from this
result via the Banks-Casher relation. With the average instanton 
density $N/V$ and sizes $\bar\rho$ commonly used for the instanton 
medium, one gets the reasonable 
\begin{equation} 
\langle\bar\psi\psi\rangle_0 = -\frac{\pi\nu(0)}{V} 
= -\frac{2N}{\kappa V} \sim \frac{1}{\bar R^2\bar\rho} 
\approx - (255 \,{\rm MeV})^3\,.  
\end{equation} 
We note that the spectrum-smearing parameter $\kappa$ is a small 
number -- approximately 100 MeV. One likewise obtains a good 
approximation to the pion decay constant, finding $f_{\pi} \simeq$ 100 
MeV, comparable to its well-known experimental value of 93 MeV\cite{DP2,DP3}.

\begin{figure}[t]
\setlength\epsfxsize{6cm}
\centerline{\epsfbox{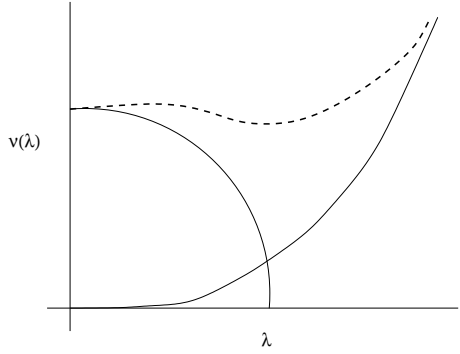}}
\caption{Schematic eigenvalue distribution of the Dirac operator.  The
solid lines are the zero mode and free contributions, the dashed line an
estimate of the full spectrum.}
\label{sdfig}
\end{figure}
There is naturally more to the spectrum than the zero modes; minimally, the
high-momentum modes which can be approximated by those of free quarks.  
The eigenvalue density obtained from this part of the spectrum is 
\begin{equation}
\nu(\lambda) \approx \frac{N_c}{4\pi^2}\lambda^3\,.
\end{equation}
The full spectrum, shown schematically in Fig. \ref{sdfig}, is a combination
of this, the Wigner semicircle, and the intermediate modes which are ignored
in this treatment.

It was noticed by Simonov\cite{Sim} that the partition function
(\ref{partfu}) can be generalized to the schematic form of a random matrix
integrated with some weight whose precise form is not
essential because of the `universality'. This allows an
economical rewriting:
\begin{equation}
{\cal Z}_{low} = \int\!dT\;P(T T^\dagger) {\rm det}\left[
\begin{array}{cc} im & T \\ T^\dagger & im \end{array} \right].
\label{simonov}
\end{equation}
With an exponential weight,
the form of eq.\,(\ref{simonov}) is the partition function of chiral random
matrix theory, from which one obtains the Wigner semicircle\cite{Sim}, see
also Ref. \cite{NVZ}. 
Later on, eq.\,(\ref{simonov}) became a base of a now well-developed
study of the microscopic spectral density of the Dirac operator and 
their correlations; see J. Verbaarschot, these proceedings.

At finite density, the expressions must naturally be modified. This 
has been done in a schematic way with random matrices\cite{HJV}.  In 
the instanton picture, we retain the microscopic physics of the zero 
mode solution at finite chemical potential, which shall be discussed in 
what follows.

\subsection{Propagator Approach}

The random matrices approach is, unfortunately, limited to the spectral
density: it cannot be used to study the $x$-dependent correlation 
functions related to chiral symmetry breaking. Knowing the microscopic
origin of the random matrices $T_{I\bar I}$ as due to the overlap 
of the instanton zero modes, allows one to calculate those quantities.

The simplest quantity is the ensemble-averaged quark propagator 
computed in refs. \cite{DP2} (second and third reference) and in 
ref. \cite{Pob}.
The propagating quark interacts with the background via the would-be zero
modes, quantified in the same overlap matrices $T_{IJ}$ used previously.
It has the averaged form:
\begin{equation}
S(x-y) = \overline{ \Phi^*_I(x-z_I)\frac{1}{im+T_{IJ}}\Phi_J(y-z_J)} =
\frac{1}{\dpp -iM(p)}\,.
\end{equation}
The result is a dynamical quark mass $M(p) = const \cdot 
\sqrt{\bar{\rho}^2 N/V} F(p\bar\rho)$, where $F(p\bar\rho)$ is a 
function known analytically and features a natural cutoff at momentum 
$p\bar\rho=1$.

Closing a quark loop with this effective propagator generates a `gap
equation' encoding the instanton density dependence of the dynamical 
mass\cite{DP2}, 
\begin{equation} 
\frac{N}{V}=4N_c\int\frac{d^4p}{(2\pi)^4}\;\frac{M(p)^2}{p^2+M(p)^2}\,.
\end{equation}
Since it has a solution with $M(p)\neq 0$, even for a zero current 
quark mass $m$, it implies spontaneous chiral symmetry breaking.
A nonzero current quark mass $m$ can be also incorporated in this
approach\cite{DP2,Pob}. Knowing the propagator $S(x,y;m)$ one can
find the determinant for low-momentum modes according to a general 
formula
\begin{eqnarray}
{\rm Det}_{low} &=& {\rm exp}\int_m^{M_1}\!dm' \left[ S(x,x;m') -
S_0(x,x;m')\right] \nonumber\\
&=& {\rm exp}\left\{-N\int_m^{M_1}\!dm' \frac{1}{2\kappa^2}
\left( \sqrt{m'^2 + 4\kappa^2}-m'\right)\right\}\,.
\end{eqnarray}
Comparing it with another representation,
\begin{equation} 
{\rm Det}_{low} = \exp\left(\frac{1}{2}\int d\lambda\,\nu(\lambda)
\ln\frac{\lambda^2+m^2}{\lambda^2+M_1^2}\right)\,,
\end{equation}
one recovers the semicircle spectrum, eq.\,(\ref{Wigner}). It 
demonstrates the mathematical equivalence of the propagator and the 
random matrices approaches, though the first method is more powerful as 
it allows one to compute also $x$-dependent correlations\cite{DP2,DP3}. 
In both cases the large-$N_c$ approximation has been exploited.

\subsection{Effective Lagrangian Approach}

In the previous section the effective quark propagator was obtained by
averaging the instanton background's effects on a single quark.  One can
recover identical physics by first averaging over instantons at the level
of the partition function.  The result is an effective action for quarks
which contains instanton effects in induced multi-quark interactions.  The
would-be zero modes serve here as a bridge, passing information from the
instanton vacuum to the effective quarks through the induced vertex.  The
consequent interactions are vertices involving $2N_f$ quarks, commonly cited
as 't Hooft interactions since he was the first to specify the 
proper quantum numbers.

The overall strength of this interaction is not fixed once and forever 
but rather has to be integrated over all possible values, since
it technically appears as a Lagrange multiplier\cite{D1}.
Fortunately, in the thermodynamic
limit one finds the integral is completely determined by its saddle-point
result.  Denoting the coupling constants $\lambda_\pm$, the result for the
low-momentum partition function is\cite{D1,CD}
\begin{eqnarray}
{\cal Z} &=& \int\!d\lambda_+d\lambda_-\int
\!D\psi\:D\psi^\dagger\:\exp\Bigg\{\int\!d^4x\:\psi^\dagger(i\dd-i\mu\gamma_4)
\psi + \lambda_+ Y^+_{N_f} + \lambda_- Y^-_{N_f}\nonumber\\
&& \quad\quad\quad\quad + N_+\left(\ln\frac{N_+}{\lambda_+ V} -1 \right) +
 N_-\left(\ln\frac{N_-}{\lambda_- V} -1 \right) \Bigg\}.
\label{Z5}
\end{eqnarray}
The left-handed vertex,
\begin{eqnarray}
Y^+[\psi,\psi^\dagger]  &=& \int\!dU\;
\int\!\prod_f^{N_f}\left[\frac{(d^4p_fd^4k_f)}{(2\pi)^{8}}\right]
(2\pi)^4\delta^4\Big(\sum(p_f-k_f)\Big)\prod^{N_f}_f \Big[
U_{l_f}^{\alpha_f} U_{\beta_f}^{\dagger o_f}
\nonumber\\
&&\cdot \psi^\dagger_{Lf\alpha_f i_f}(p_f)
{\cal F}(p_f,\mu)_{k_f}^{i_f}\epsilon^{k_fl_f} \epsilon_{n_fo_f}
{\cal F}^\dagger(k_f,-\mu)_{p_f}^{n_f}\psi_L^{f\beta_fp_f}(k_f)\Big]\,
\label{vertex}
\end{eqnarray}
contains the matrix form factors
\begin{equation}
{\cal F}(p,\mu) = (p+i\mu)^- \varphi(p,\mu)^+\, , \quad\;\;\;
{\cal F}^\dagger(p,-\mu) = \varphi^*(p,-\mu)^-(p+i\mu)^+\,.
\end{equation}
The $\varphi(p,\mu)$ are the Fourier-transformed zero mode solutions and we
use the notation notation $x^{\pm}=x^\mu\sigma_\mu^{\pm}$, where the $2
\times 2$ matrices $\sigma_\mu^{\pm} = (\pm i \vec \sigma,1)$ decompose
the Dirac matrices into chiral components,
and it is understood that $\mu$ written as a four-vector is
$\mu_\alpha = (\vec 0,\mu)$.  A similarly defined $Y^-$ carries right-handed
quarks.  All calculations discussed here are in the topologically neutral
case, where $N_+=N_-=N$ and hence $\lambda=\lambda_+=\lambda_-$.

With a Fierz decomposition, the instanton-induced interaction can easily be
made to resemble the Nambu--Jona-Lasinio model. However, in the steps
leading to this result, microscopic information has been retained. Instantons
not only provide a natural cutoff through the form factors and a dynamically
determined coupling strength, but also account for anomalously broken
U$_A(1)$ symmetry and correctly reproduce SU(4) Pauli-G\"ursey symmetry
when $N_c=2$. It should be also added that the naive addition of a 
nonzero current quark mass to the NJL Lagrangian fails to reproduce
several known low-energy Ward identities, as well as the 
phenomenologically-known coefficients in the Gasser--Leutwyler
chiral Lagrangian (the terms containing $m^2$ and $m\cdot p^2$).
The microscopic instanton approach preserving all symmetries
of QCD is capable to correctly incorporate nonzero quark masses, 
and it does so in a rather nontrivial way\cite{Mus}.

\section{Competition Between $\bar{q}q$ and $qq$ Channels}

Since the instanton-induced interactions (\ref{vertex}) support both
$\bar q q $ and $q q $ condensation, it is necessary to consider the two
competing channels simultaneously. This means that one must calculate
both the normal ($S$) and anomalous ($F$) quark Green functions. A
colour/flavour/spin ansatz compatible with the possibility of chiral
and colour symmetry breaking is
\begin{eqnarray}
\langle\psi^{f\alpha i}(p)\psi^\dagger_{g\beta j}(p)\rangle &=&
\delta^f_g \delta^\alpha_\beta S_1(p)^i_j \quad
{\rm for}\:\alpha,\beta=1,2 \,,
\nonumber
\\
\langle\psi^{f\alpha i}(p)\psi^\dagger_{g\beta j}(p)\rangle &=&
\delta^f_g \delta^\alpha_\beta S_2(p)^i_j \quad
{\rm for}\:\alpha,\beta>2 \,,
\nonumber
\\
\langle\psi_L^{f\alpha i}(p)\psi_L^{g\beta j}(-p)\rangle
&=&
\langle\psi_R^{f\alpha i}(p)\psi_R^{g\beta j}(-p)\rangle
= \epsilon^{fg}\epsilon^{\alpha\beta[\gamma]}\epsilon^{ij} F(p)\,,
\end{eqnarray}
where $[\gamma]$ refers to some generalized direction(s) in colour space,
and it is this set of $N_c-2$ indices which signals the breaking of
colour symmetry.  In the particular case of $N_c=3$, where the colour
symmetry is broken as $SU(3)\rightarrow SU(2)\times U(1)$ and our
ansatz considers the $\bar{3}$ channel, we will by
convention take $[\gamma]=3$; for $N_c=4$ one can take
$[\gamma]=34$ and so forth. In the event of colour symmetry
breaking, the standard propagators (and ensuing condensates) will lose
their colour degeneracy and the separation of $S(p)$ into $S_1(p)$ and
$S_2(p)$ becomes necessary; otherwise the Schwinger-Dyson-Gorkov equations 
do not close\cite{CD}.

Written in the chiral $L,R$ basis, the $4\times 4$ propagators
$S_{1,2}(p)$ are of the form:
\begin{equation}
S(p) = \left[\begin{array}{cc}G(p) {\bf 1} & Z(p){\bf S}_0(p)^+ \\
Z(p){\bf S}_0(p)^- & G(p) {\bf 1} \end{array}\right] \,.
\end{equation}
Here the off-diagonal, bare propagator
${\bf S}_0(p)^{\pm} = \left[(p+i\mu)^\mp\right]^{-1}$
is modified by the scalar functions $Z_{1,2}(p)$, and is augmented on the
diagonal by the scalar $G_{1,2}(p)$ which if nonzero break chiral symmetry.

\begin{figure}[bt]
\setlength\epsfxsize{10cm}
\centerline{\epsfbox{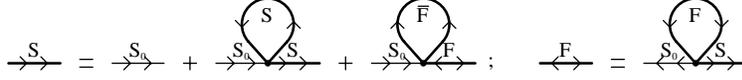}}
\caption{Schwinger-Dyson-Gorkov diagrams to first order in $\lambda$.}
\label{gorkfig}
\end{figure}

Using the instanton-induced interaction (\ref{vertex}) directly, without
Fierz rearrangement, one can build a
systematic expansion for the $F,G$ Green functions in the parameters $1/N_c$
and $\bar\rho/\bar R$. In the leading order in both
parameters we restrict ourselves to the one-loop approximation
shown in Fig. \ref{gorkfig}. It corresponds to a set of self-consistent
Schwinger-Dyson-Gorkov equations.
An important $\mu$-dependence enters through the form factors in 
eq.\,(\ref{vertex}).

\subsection{The System of Gap Equations}

This system of equations reduces to a set of three gap equations which
self-consistently specify three condensates:  two chiral and one diquark.
They are defined as the closed loops:
\begin{equation}
g_{1,2} = \frac{\lambda}{N_c^2-1} \int\!\frac{d^4k}{(2\pi)^4}\:
A(k,\mu)G_{1,2}(k) \,;\quad
f = \frac{\lambda}{N_c^2-1} \int\!\frac{d^4k}{(2\pi)^4}\: B(k,\mu)F(k) \,.
\label{condefs}
\end{equation}
The functions $A(k,\mu)$ and $B(k,\mu)$ are scalar functions which arise
from spin-averaging the matrix form factors, as defined in ref. \cite{CD}.
The quantities $M_{1,2}$ and $\Delta$ are linear
combinations of the condensates $g_{1,2}$ and $f$:
\begin{eqnarray}
M_1 &=& \left(5-4/N_c\right)g_1+ \left(2 N_c-5+2/N_c\right)g_2
\, , \nonumber\\
M_2 &=& 2\left(2-1/N_c\right)g_1+2(N_c-2)g_2\,,\quad
\Delta = \left(1+1/N_c\right)f\,.
\label{defs}
\end{eqnarray}
The $M_{1,2}$ are measures of chiral symmetry breaking, which
act as an effective mass modifying the standard quark propagation.  On the
other hand the diquark loop $2\Delta$ plays the role of twice
the single-quark energy gap formed around the Fermi surface.

The solution of the gap equations depends on the vertex coupling constant,
$\lambda$, which itself is determined by balancing the instanton background
with the condensates through its saddle-point value.  This minimization of
the partition function (\ref{Z5}) leads to
\begin{equation}
\frac{N}{V} = \lambda\langle Y^+ + Y^-\rangle = \frac{4(N_c^2-1)}{\lambda}
\left[ 2 g_1 M_1 + (N_c-2) g_2M_2 + 4 f\Delta \right]\,.
\label{gapeqn}
\end{equation}
This joins the gap equations to close a system of equations, numerically
solvable.

Once this is done, the chiral condensate proper may be computed as an
integral over the resummed propagator:
\begin{equation}
 -\langle\bar{\psi}\psi\rangle_{Mink} =
i\langle\psi^\dagger\psi \rangle_{Eucl} = 4\int\!
\frac{d^4p}{(2\pi)^4}\,\left[2 G_1(p) + (N_c-2) G_2(p)\right] .
\end{equation}

\subsection{Thermodynamic Competition}

For any given chemical potential, multiple solutions can be obtained for
the gaps.  These correspond to different phases of quark matter, and they
are summarized as follows:

{\it (0)  Free massless quarks:} $g_1=g_2=f=0$.

{\it (1)  Pure chiral symmetry breaking:} $g_1=g_2\ne 0$, $f=0$.  

{\it (2)  Pure diquark condensation:} $g_1=g_2=0$, $f\ne 0$.  

{\it (3)  Mixed symmetry breaking:}  $g_1 \ne g_2 \ne 0$, $f\ne 0$.

The free energy, calculated to first order in $\lambda$, is minimized 
in order to resolve the stable solution.
The phase corresponding to the {\em lowest} coupling $\lambda$
is the thermodynamically favoured\cite{CD}.

\begin{figure}[t]
\leavevmode
\begin{center}
\setlength\epsfxsize{5.5cm}
\epsfbox{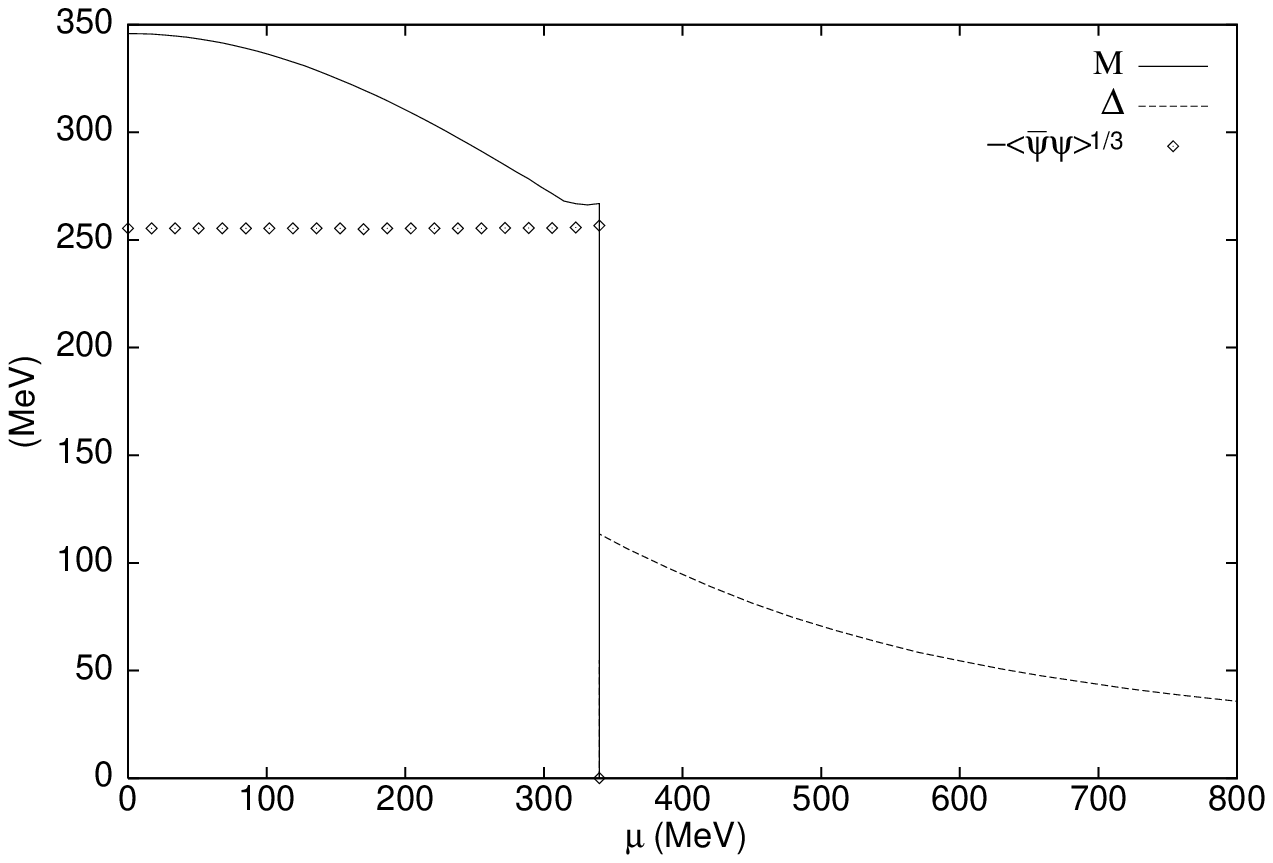}
\setlength\epsfxsize{5.5cm}
\epsfbox{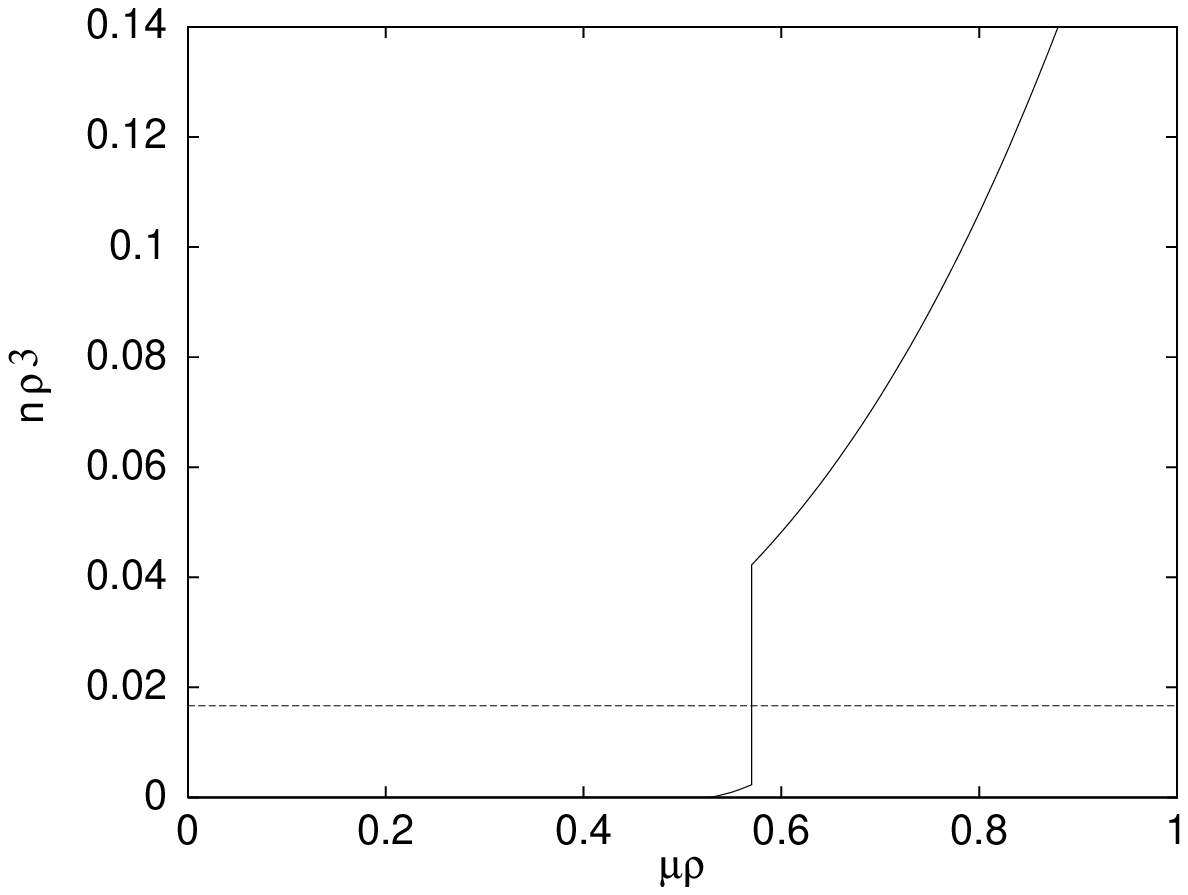}
\end{center}
\caption{Left panel: Condensates for $N_c=3$ as a function of $\mu$.
Shown are the effective quark mass $M$, the quark condensate
$-\langle\bar\psi\psi\rangle^{1/3}$, and the diquark energy gap per
quark $\Delta$.  Right Panel:  The quark density $n_q$ vs. $\mu$.}
\label{phasedia}
\end{figure}

No solutions were found matching Phase (0), and the Phase (3) solution
obtained disappears at relatively low chemical potential ($\mu \approx
80$ MeV) and is never thermodynamically competitive\cite{CD}.  The remaining
phase competition is then between Phases (1) and (2).  In the vacuum, where
$\mu =0$, one finds Phase (1) preferred -- this is the standard picture.
However, at a critical chemical potential $\mu_c$, defined by the ratio
of superconductive gap to chiral effective masses
\begin{equation}
\frac{\Delta(\mu_c)}{M(\mu_c)} = \sqrt{\frac{N_c}{8(N_c-1)}} =
\frac{\sqrt{3}}{4} \,,
\end{equation}
a first-order phase transition occurs.  With the standard instanton parameters
$N/V = 1$ fm$^{-4}$ and $\bar\rho=0.33$ fm, we find $\mu_c\simeq 340$ MeV.
This and the first-order nature of the phase transition are clearly seen
in Fig. \ref{phasedia}.

\begin{figure}[bt]
\begin{center}
\leavevmode
\setlength\epsfxsize{5.5cm}
\epsfbox{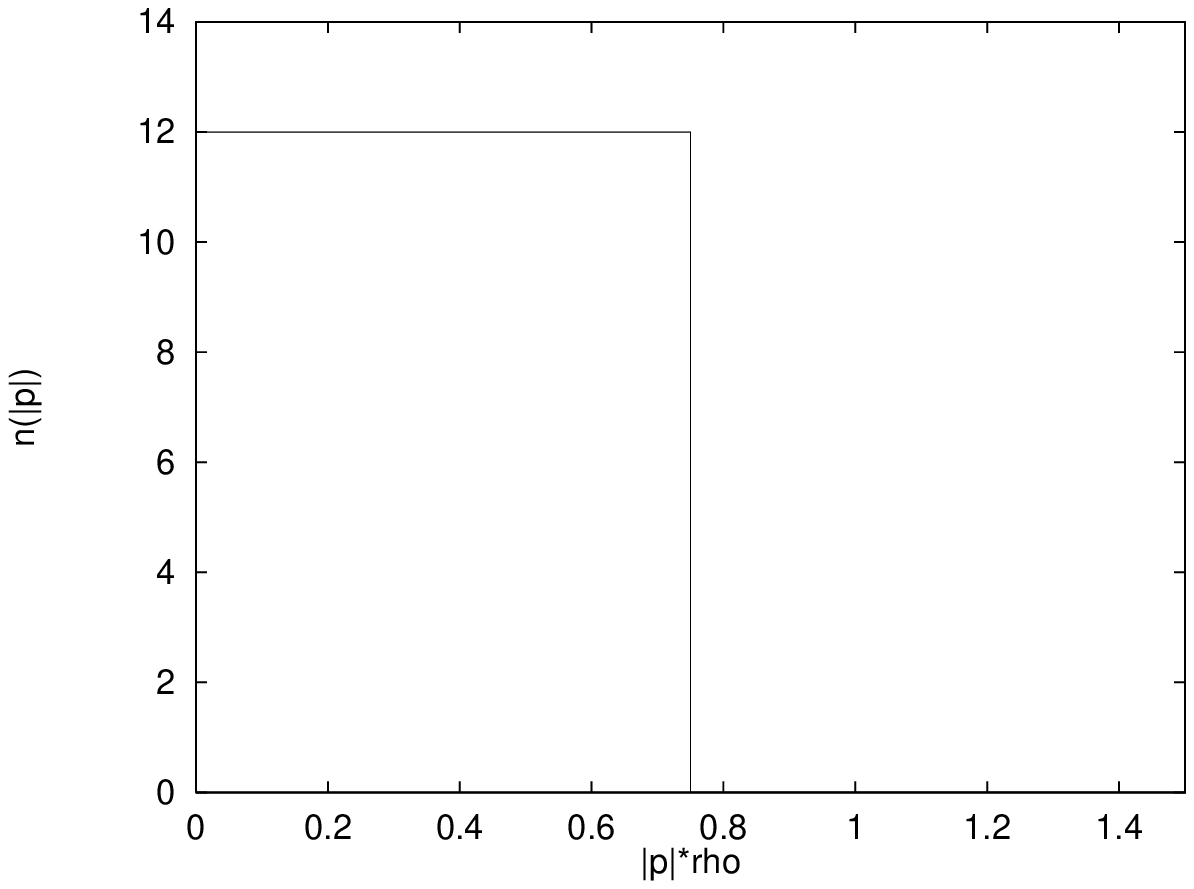}
\setlength\epsfxsize{5.5cm}
\epsfbox{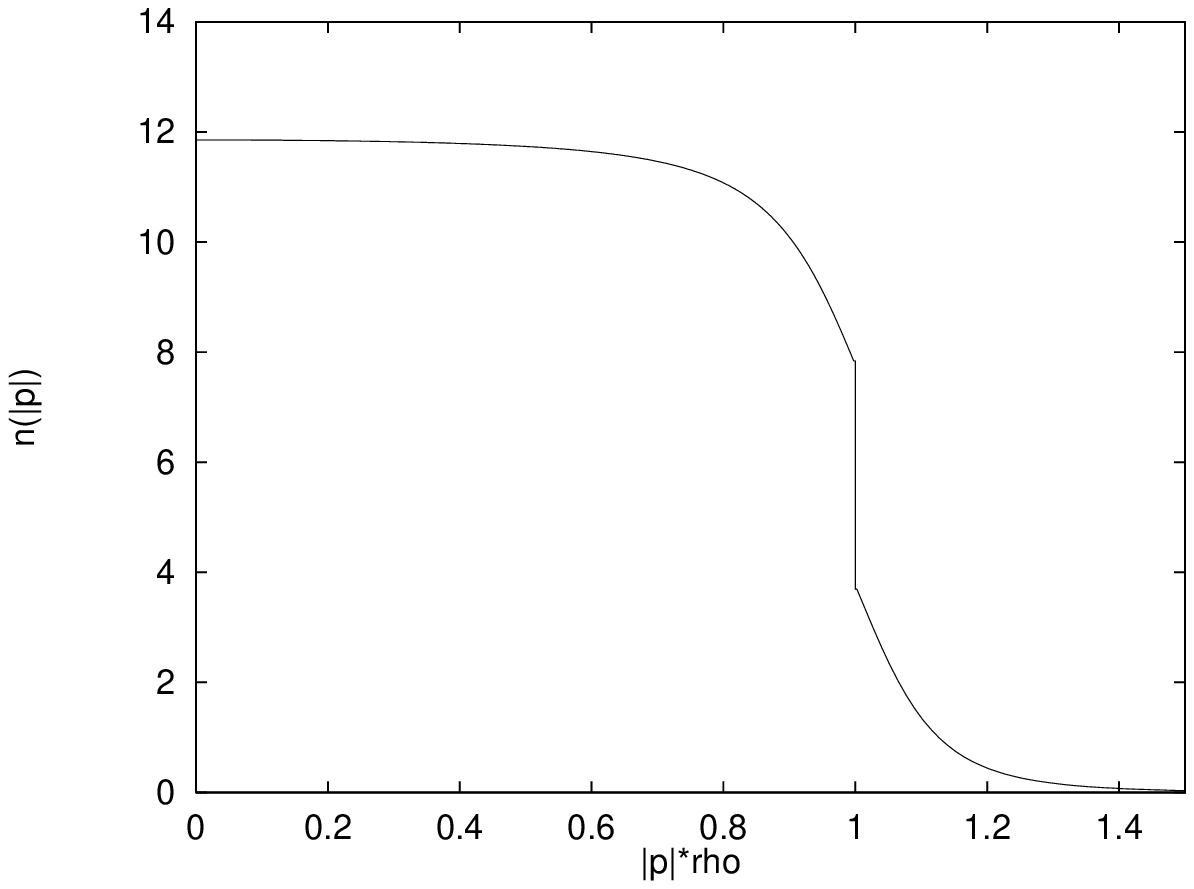}
\end{center}
\caption{Occupation number $n(p)$ vs. $p$ for Phase (1) (left panel) and
Phase (2) (right panel) for $\mu=1/\bar\rho=600$ MeV.}
\label{occfig}
\end{figure}

The quark density is physically more relevant than the chemical 
potential.  Thus we calculate
\begin{equation}
n_q = \int\,d^4x\,j_4(x) = -i \int \frac{d^3p}{(2\pi)^3} \; n(p)\,,
\end{equation}
where we have defined the occupation number density for quarks as $n(|\vec{p}
|)$ as an intermediate step.  The correct, conserved form of the quark
current is given in ref. \cite{CD}.  Here we present
only numerical results, in Figs. \ref{phasedia} and \ref{occfig}.  
In Phase (1), there is clearly
an effective mass brought about by spontaneous symmetry breaking, indicated
by the reduced Fermi radius.  We stress that, despite the complicated 
four-momentum dependence of the interaction, the
resulting occupation number density appears as a perfect Fermi step
function.  Cooper pairing, however, smears the Fermi
surface, and this is evinced in the second plot.  The residual discontinuity
at $|\vec{p}|=\mu$ is the contribution from the free, colour-3 quarks
which do not participate in the diquark.  Those that do are smoothly
distributed as a Bosonic condensate.

Performing the integration over three-momenta, we arrive at the density as
a function of chemical potential.  In the right panel of Fig.
\ref{phasedia} this is shown for the equilibria states, demonstrating the large
discontinuity at the phase transition, where the horizontal line signifies
the quark density of equilibrium nuclear matter.  The phase transition
occurs at an extremely low quark density, a conceptual conundrum which is
seeing ongoing discussion in the literature\cite{CD,ARW,BR,Bu,BJW}.

\section{Conclusions}

We have discussed three ways of obtaining the spontaneous breaking of chiral
symmetry in the instanton approach.  While the physics of each seem
superficially distinct, it has been demonstrated that they are equivalent.
One of these approaches was extended for and applied to finite density quark
matter, where a chiral symmetry restoring, first order phase transition
reorders the quarks into a colour superconducting medium.  These results
agree with related works, but unfortunately also share a mysteriously low
transition density, a mere fraction of that corresponding to equilibrium
nuclear matter.

\end{document}